\documentclass[nonacm, sigconf]{acmart}

\usepackage{comment}
\usepackage{csquotes}
\usepackage{xspace}
\usepackage{enumitem}
\usepackage{tabularx, booktabs, multirow, makecell}
\usepackage{listings}
\usepackage{setspace}
\usepackage{subcaption}
\usepackage{graphicx}
\usepackage{tikz}
\usepackage{tablefootnote} 

\AtBeginDocument{
  }

\acmISBN{978-1-4503-XXXX-X/2018/06}

\begin{document}

\newcommand\todoit[1]{{\color{red}\{TODO: \textit{#1}\}}}
\newcommand\todocite{{\color{red}{CITE}}}

\newcommand\donghoon[1]{{\color{blue}\textit{\textbf{[DS] #1}}}}
\newcommand\emily[1]{{\color{red}\textit{\textbf{[ET] #1}}}}
\newcommand\alice[1]{{\color{purple}\textit{\textbf{[AG] #1}}}}
\newcommand\rock[1]{{\color{teal}\textit{\textbf{[RP] #1}}}}
\newcommand\jae[1]{{\color{orange}\textit{\textbf{[JL] #1}}}}

\definecolor{lightblue}{RGB}{212, 235, 255}
\definecolor{orange}{RGB}{255, 105, 0}
\definecolor{lightgreen}{RGB}{177, 231, 171}
\definecolor{lightyellow}{RGB}{255, 255, 148}

\definecolor{colorRepresentational}{HTML}{E41A1B}
\definecolor{colorQuality}{HTML}{377EB8}
\definecolor{colorCognitive}{HTML}{4EAF4A}
\definecolor{colorAgency}{HTML}{984EA4}
\definecolor{colorAllocative}{HTML}{FF7F00}
\definecolor{colorTechnical}{HTML}{E3E33F}
\definecolor{colorInterpersonal}{HTML}{A75629}

\newcommand\tworows[1]{\multirow{2}{*}{\shortstack[l]{#1}}}
\newcommand\tworowsc[1]{\multirow{2}{*}{\shortstack[c]{#1}}}
\newcommand\threerows[1]{\multirow{3}{*}{\shortstack[l]{#1}}}
\newcommand{\leftcell}[2][l]{%
  \begin{tabular}[#1]{@{}l@{}}#2\end{tabular}}
\newcommand{\midcell}[2][l]{%
  \begin{tabular}[#1]{@{}c@{}}#2\end{tabular}}

\newcommand{\eg}{\textit{e.g.}}
\newcommand{\ie}{\textit{i.e.}}
\newcommand{\cf}{\textit{c.f.}}
\newcommand{\etal}{\textit{et al.}}

\newcolumntype{P}[1]{>{\centering\arraybackslash}p{#1}}
\newcolumntype{L}[1]{>{\raggedright\let\newline\\\arraybackslash\hspace{0pt}}m{#1}}
\newcolumntype{R}[1]{>{\raggedleft\arraybackslash}p{#1}}

\definecolor{mypurple}{HTML}{9966CC}

\newcommand{\purpledot}[1]{%
  \tikz[baseline=(char.base)]{
    \node[shape=circle,draw=none,fill=mypurple,
      text=white,inner sep=1pt,font=\bfseries] (char) {#1};
  }%
}

\newcommand{\sampleCountry}{{\textit{country}}\xspace}

\newcommand{\sampleCategory}{{\textit{topic}}\xspace}

\title{Interrogating Design Homogenization in Web Vibe Coding}

\author{Donghoon Shin}
\orcid{0000-0001-9689-7841}
\authornote{Work done during an internship at Microsoft Research.}
\affiliation{%
  \institution{University of Washington}
  \city{Seattle}
  \state{WA}
  \country{USA}}
  \email{dhoon@uw.edu}
  
\author{Alice Gao}
\affiliation{%
  \institution{University of Washington}
  \city{Seattle}
  \state{WA}
  \country{USA}}
  \email{atgao@cs.washington.edu}
  
\author{Rock Yuren Pang}
\orcid{0000-0001-8613-498X}
\affiliation{%
  \institution{University of Washington}
  \city{Seattle}
  \state{WA}
  \country{USA}}
  \email{ypang2@cs.washington.edu}
  
\author{Jaewook Lee}
\orcid{0000-0002-1481-9290}
\affiliation{%
  \institution{University of Washington}
  \city{Seattle}
  \state{WA}
  \country{USA}}
  \email{jaewook4@cs.washington.edu}
  
\author{Katharina Reinecke}
\orcid{0000-0001-7897-9325}
\affiliation{%
  \institution{University of Washington}
  \city{Seattle}
  \state{WA}
  \country{USA}}
  \email{reinecke@cs.washington.edu}
  
\author{Emily Tseng}
\orcid{0000-0001-9689-7841}
\affiliation{%
  \institution{Microsoft Research \&\\University of Washington}
  \city{Seattle}
  \state{WA}
  \country{USA}}
  \email{emtseng@uw.edu}

\renewcommand{\shortauthors}{Shin, et al.}

\begin{abstract}
Generative AI is known for its tendency to homogenize, often reproducing dominant style conventions found in training data. However, it remains unclear how these homogenizing effects extend to complex structural tasks like web design. As lay creators increasingly turn to LLMs to `vibe-code' websites---prompting for aesthetic and functional goals rather than writing code---they may inadvertently narrow the diversity of their designs, and limit creative expression throughout the internet. In this paper, we interrogate the possibility of design homogenization in web vibe coding. We first characterize the vibe coding lifecycle, pinpointing stages where homogenization risks may arise. We then conduct a sociotechnical risk analysis unpacking the potential harms of web vibe coding and their interaction with design homogenization. We identify that the push for frictionless generation can exacerbate homogenization and its harms. Finally, we propose a mitigation framework centered on the idea of \textit{productive friction}. Through case studies at the micro, meso, and macro levels, we show how centering productive friction can empower creators to challenge default outputs and preserve diverse expression in AI-mediated web design.
\end{abstract}

\maketitle

\section{Introduction}

The rise of generative AI has transformed creative expression, enabling a broader range of individuals, or \textit{lay creators}, to craft digital artifacts without technical expertise. This shift has given rise to the emerging practice of \textit{vibe coding}, where creators use high-level, informal prompts to convey a desired \textit{vibe}~\cite{sarkar2025vibe, sapkota2025vibe}, relying on large language models (LLMs) to interpret these abstractions into functional websites. This workflow promises to democratize web development by leveraging LLMs' capabilities in rapid prototyping and aesthetic interpretation~\cite{sarkar2025vibe}. However, this convenience comes with a trade-off; by offloading structural and aesthetic decisions to AI models, lay creators increasingly depend on the latent patterns embedded within the AI to fill in their creative gaps.

This reliance offers a new and urgent site of inquiry. The spread of web vibe coding means generative AI is increasingly used to construct the web interfaces through which people connect, work, transact, and live; however, those same AI systems are known for their tendency to reproduce dominant style conventions found in training data~\cite{agarwal2025ai}. Primarily trained on English-centric datasets~\cite{johnson2022ghost, brown2020language}, such models often align with Western aesthetic conventions~\cite{oppenlaender2023perceptions, anderson2024homogenization, johnson2022ghost, liu2023cultural}. While these biases are well-documented in domains such as text~\cite{santy2023nlpositionality, ye2025semantic} and image generation~\cite{nyaaba2024generative, liu2025salt, qadri2023ai}, it is unclear how they manifest in interactive creative domains like web development. Richness in creative expression relies on the creator's diverse perspectives~\cite{wang2025adaptive}, and replacing intentional choices with probabilistic defaults risks accelerating design homogenization---a convergence toward the mean, dominant aesthetic~\cite{solaiman2023evaluating, castro2023human}. We posit that vibe coding may accelerate design homogenization by creating a direct pathway from a model's latent biases to finished creative work, especially when the mechanisms of generation prioritize speed over distinctiveness.

In this paper, we interrogate design homogenization in web vibe coding, unpacking its conceptual underpinnings to anticipate and potentially mitigate its risks. We first characterize the vibe coding lifecycle by systematically reviewing academic and gray literature on the practice, and conducting walkthroughs of popular vibe coding tools. Our lifecycle breaks down vibe coding into stages---from user prompting to deployment---and identifies where homogenization risks may arise. Then, building on this lifecycle, we conduct a sociotechnical risk analysis adapting existing harm frameworks (\ie{},~\cite{suresh2021framework, weidinger2022taxonomy, buccinca2023aha, shelby2023sociotechnical}) to web vibe coding. Our analysis finds that LLMs' tendencies towards homogenization work in tandem with mounting social pressures in the interaction design industry to create a sociotechnical environment encouraging creators to accept the path of least resistance. Risks, such as cognitive overload and overreliance on the vibe-coded design outputs, incentivize lay creators to accept the model's default output rather than iterating on a diverse and creative design.

We conclude by translating these insights into a mitigation strategy that challenges the paradigm of frictionless design~\cite{s1, Ganci2017OnWBA, acharya2025generative, fawzy2025vibe}. We argue that addressing the potential homogenization requires introducing \textit{productive friction} into the vibe coding workflow. Moving away from a frictionless paradigm where LLMs execute plans automatically, we propose frameworks where the model acts as a mediator, explicitly querying design assumptions and presenting alternatives before finalizing output. We structure these across three layers: micro (individual creator), meso (organization), and macro (ecosystem). By shifting the goal from immediate output to interactive co-creation, we outline a research agenda for designing tools that empower creators to maintain diverse, authentic web aesthetics against the pressure of automation.
We close by unpacking where, within this agenda, the communities around responsible technology can leverage its strengths to work in collaboration with allied communities in HCI and AI, in the subfield of human-AI interaction.

In summary, this paper contributes:
\begin{itemize}[leftmargin=20pt, topsep=0pt, partopsep=0pt, parsep=0pt, itemsep=0pt]
\item A systematic review of academic and gray literature around vibe coding, resulting in a comprehensive characterization of the vibe coding lifecycle (\S\ref{sec:step1}), pinpointing stages where potential homogenization risks may intrude;
\item A risk analysis (\S\ref{sec:step2}) unpacking harms in vibe coding and their intersection with design homogenization, and identifying frictionless generation as an active contributor to homogenization;
\item A multi-level mitigation framework (\S\ref{sec:step3}) that proposes \textit{productive friction} as a design intervention, offering case studies at the micro, meso, and macro levels, and offers a collaborative research agenda for teams of scholars across human-AI interaction.
\end{itemize}
\section{Related Work}

\subsection{Harms and Homogenization in Generative AI}

In HCI, AI, and allied communities around responsible technology, the widespread consumer release of LLMs has driven a surge of work examining the societal risks and harms of generative AI. Open-ended interactions with current AI systems may shape perception, behavior, and social norms in ways that the stakeholders must understand to guide these systems toward positive ends~\cite{gautam2024melting, vassel2024psychosocial, ferrara2024genai}. As such, prior research has sought to taxonomize the range of generative AI harms. For example, Weidinger \etal{}~\cite{weidinger2022taxonomy} developed a taxonomy of ethical and social risks associated with language models, identifying six categories (\ie{}, discrimination, information hazards, misinformation, malicious uses, HCI harms, and environmental harms). Suresh \& Guttag~\cite{suresh2021framework} examined harms across the full machine learning (ML) lifecycle to anticipate, prevent, and mitigate downstream effects. Identifying issues like historical, representation, and deployment bias, they emphasized that harms do not stem solely from data, but can arise at multiple stages of the ML pipeline, such as data collection, preparation, model development, evaluation, postprocessing, and deployment.

In response, several recent studies have built on these broader characterizations of generative AI harms through in-depth studies in specific applications, thus revealing additional severe and persistent problems. For instance, Agarwal \etal{}~\cite{agarwal2025ai} found that LLM writing assistants homogenize users' output toward Western-preferred styles, thus diminishing cultural nuances. This homogenizing effect may compound with the risks on long-term creativity, as Kumar \etal{}~\cite{kumar2025human} revealed that, while LLM assistance can provide short-term boosts in creativity during assisted tasks, it may inadvertently hinder independent creative performance when users work without assistance. Similar risks to short- and long-term creativity may also manifest in vibe coding: studies have found developers may over-rely on AI-generated artifacts in programming tasks~\cite{pearce2025asleep}. This risks premature creative closure, as users may uncritically accept model suggestions rather than exploring the divergent possibilities for creative expression. These risks of homogenization, creative deskilling, and over-reliance threaten to become deeply embedded in emerging practices like vibe coding, particularly for lay individuals who may lack the technical expertise to critically evaluate or diverge from AI output.

In this paper, we conceptualize vibe coding as a sociotechnical practice, detailing the risks for lay creators throughout its end-to-end lifecycle. We address the unique harms faced by creative intermediaries who lack technical expertise yet significantly influence web design outcomes for broader audiences. In short, we build on the methodological foundation of taxonomizing general risks in LLMs, but focus on risks from an emerging practice---vibe coding---for lay creators.

\begin{figure*}[t!]
    \centering
    \includegraphics[width=.7\linewidth]{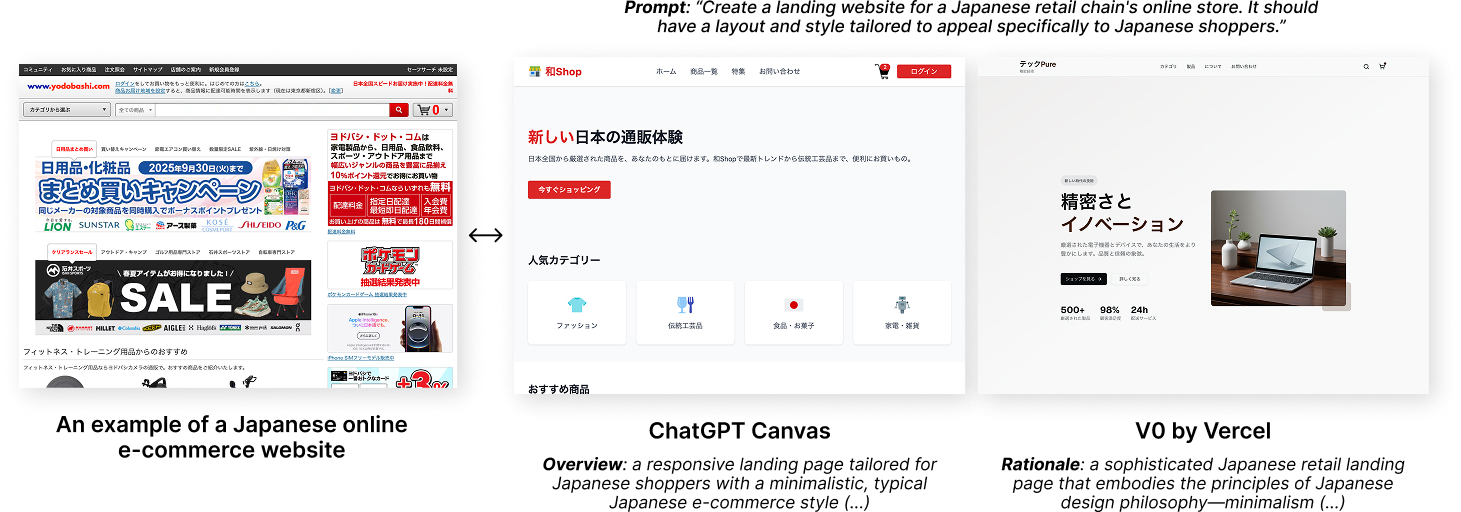}
    \caption{Example of design homogenization where global defaults override regional preference. The models justified a minimalist layout (right) as a cultural preference, thereby overriding the preference for high information density in Japanese web design~\cite{baughen2021patterns, nordhoff2018case}.}
    \label{fig:reactive_example}
\end{figure*}

\subsection{AI in Coding Support \& Vibe Coding}

AI has long supported programmers' work, from traditional tools for syntax highlighting and templated code completion to workflow assistants embedded in IDEs supporting documentation, refactoring, and debugging~\cite{mastropaolo2023robustness}. With the advent of generative AI, coding support has become significantly more contextual and powerful. LLMs (\eg{}, GPT) and their fine-tuned variants---trained on vast amounts of code---have shown promising performance in natural language processing and applied to assist in code writing for programmers (\eg{}, code completion)~\cite{brown2020language, fan2023automated}. This has led to the development of sophisticated AI programming assistants (\eg{}, GitHub Copilot), which represent a new concept of `AI pair programmers'~\cite{mcnutt2023design, chen2021evaluating}. Similarly, editor-integrated AI tools (\eg{}, Cursor, Windsurf) let programmers describe desired changes in natural language on the IDE and have it autocomplete a snippet or refactor code.\footnote{Although occasionally referred to as vibe coding tools, these tools operate within an IDE and therefore may not be accessible to non-technical creators. Given our focus, we center our analysis on tools operable without coding expertise.}

Despite these, even most AI-supported programmers continue to need programming knowledge to guide, debug, and integrate AI-generated code effectively~\cite{zviel2024good, eibl2025exploring}. In this context, no-code vibe coding emerges as a promising approach, enabling individuals without coding knowledge to create artifacts by describing desired outcomes or vibes in natural language, which AI interprets to generate functional code~\cite{sarkar2025vibe, eibl2025exploring, sergeyuk2025using}. This sets the approach apart from other editor-based web builders (\eg{}, WYSIWYG editors) by shifting the primary interaction from manual manipulation to natural language conversation. As such, it is often referred to as democratization of software development---particularly for website creation by lay users~\cite{geng2025exploring}, because it inherently involves intuitive, high-level concepts like layout, aesthetics, interactivity, and user experience that can be vividly described in everyday language, allowing AI to bridge the gap between vague ideas and deployable code without requiring extensive technical expertise. This aligns with the no-code paradigm, reducing barriers and enabling creators to concentrate on high-level concepts rather than syntactic details.

To date, however, few works have explored the impacts of vibe coding, particularly its potential harms for lay creators. Since individuals with little coding knowledge may heavily rely on AI outputs, they are more vulnerable to biases and risks embedded in the generated code, such as security vulnerabilities and discriminatory patterns inherited from biased training data~\cite{pearce2025asleep, nguyen2022empirical}. For instance, AI-generated code can introduce common weaknesses like out-of-bounds writes, potentially leading to exploitable flaws in applications developed by lay users~\cite{pearce2025asleep}. Moreover, algorithmic biases can perpetuate societal prejudices, such as gender or racial discrimination (\eg{},~\cite{ren2024survey, urchs2025all, urman2024foreign}), risking to amplify harms in user-facing websites and reduce trust in AI-assisted development. These issues highlight the necessity to investigate the harms caused by AI in vibe coding processes like developing mitigation strategies---including, but not limited to, bias-aware datasets and user guidance---to ensure safe and equitable use for lay creators in website development.

\section{Step 1: Characterizing the End-to-End Lifecycle of Vibe Coding for Web Creation}\label{sec:step1}

To identify where design homogenization may emerge and where interventions could be introduced, we began by characterizing the end-to-end lifecycle of vibe coding a website.
Doing so allowed us to break down vibe coding into distinct stages, enabling us to further analyze where risks of design homogenization may arise, and where the industrial dynamics of interface design encourage lay creators to accept default outputs.

\subsection{Methods}

As an emerging practice, much of our knowledge of vibe coding lies outside traditional academic venues. To account for this, we combined a content analysis of academic and grey literature with an analysis of contemporary tools, which enabled us to examine both how vibe coding is conceptualized and how it operates in practice.

\subsubsection{Systematic review of academic and grey literature}
\label{sec:corpus}

To examine emergent practices, user experiences, and technical discussions around vibe coding, we conducted a broad content analysis of relevant literature. We began by surveying academic databases (\ie{}, Google Scholar) for research on vibe coding. Given the novelty of the term, however, we found few academic sources directly addressing the phenomenon. To fill this gap, we expanded our review to include grey literature~\cite{rothstein2009grey}, which would more closely reflect how online users encounter and discuss vibe coding. This encompassed technical blog posts (\eg{}, Medium articles), forum discussions (\eg{}, Reddit threads), and thought leadership (\eg{}, social media posts), which offered timely, practical insights into user motivations, colloquial language, common workflows, and frustrations that have not yet been captured in academic research.

\begin{table*}[]
\caption{Comparison of example LLM services supporting web vibe coding and their observable processes for creating a website, following a user prompt. The term \textit{artifact} refers to the rendered output website produced by the system.}
\footnotesize
\begin{tabular}{l|l|l}
\toprule
\textbf{Type} & \textbf{Service} & \textbf{Process (after user prompt)}\\
\toprule
\multirow{3}{*}{General-purpose LLM service}& ChatGPT Canvas~\cite{chatgpt_canvas} & Artifact → One-line overview of the generated website \\
 & Gemini Canvas~\cite{gemini_canvas} & One-line planning → Artifact → One-line overview of the generated website \\
 & Claude Artifacts~\cite{claude_artifacts} & One-line planning → Artifact → Multi-line description of the generated website \\
\midrule
\multirow{4}{*}{Specialized vibe coding tool} & Lovable~\cite{lovable} & Multi-line planning → Artifact → Multi-line description of the generated website \\
 & v0 by Vercel~\cite{v0} & Artifact → Multi-line description of the generated website\\
 & Replit~\cite{replit}& \leftcell{Multi-line planning (spec request if prompt is too short) → Artifact\\→ Multi-line description of the generated website} \\
\bottomrule
\end{tabular}
\label{tab:tool_exploration}
\end{table*}

To systematically gather these sources, we compiled a list of keywords and programmatically retrieved results via the Google Search API.
Our search strategy for both resources was organized around three thematic categories of keyword queries (see Appendix~\ref{apdx:query} for the full list): (1) informal and vibe-centered phrases (\eg{}, \textit{`vibe coding'}); (2) interaction and technique-centered phrases (\eg{}, \textit{`prompt-based website coding'}); and (3) platform-specific use cases (\eg{}, \textit{`ChatGPT building a website'}). To filter the results, we screened these to ensure their relevance, applying a key inclusion criterion: each source must discuss vibe coding as the creation of a software artifact through text-based interaction. We also excluded sources aimed exclusively at expert programmers for workflow streamlining or those exclusively focused on non-website mediums. For academic literature, this screening process narrowed an initial pool of results down to 13 papers; for the more voluminous grey literature, we reviewed search results sequentially by initially examining 50 sources and then adding any that met our criteria to our corpus until we reached a total of 50. This dual-collection strategy resulted in our final corpus of 63 sources (13 academic and 50 grey literature) for analysis.

\subsubsection{Analysis of vibe coding platforms and tools for web creation}
Finally, to ground our lifecycle model in current available technologies, we conducted methodical walkthroughs of popular and commercially available interactive systems that facilitate vibe coding. We specifically explored six widely-used vibe coding tools, ranging from vibe coding support provided within the general-purpose LLM services (\eg{}, ChatGPT Canvas~\cite{chatgpt_canvas}, Gemini Canvas~\cite{gemini_canvas}, Claude Artifacts~\cite{claude_artifacts}) to specialized vibe coding platforms (\eg{}, Lovable~\cite{lovable}, v0 by Vercel~\cite{v0}, Replit~\cite{replit}), that support web creation. Although we sought to capture a broad set of currently popular tools, we recognize that many other smaller services also exist, as vibe coding for websites remains a rapidly evolving field.

The walkthroughs were conducted by the leading author over one week, interacting with each platform to create a website with free-form prompting while documenting workflows, user interactions, and system outputs. These walkthrough records were then collaboratively reviewed by three authors to identify recurring interaction patterns, categorize the nature of generated outputs (\eg{}, code, live previews), and examine mechanisms for refinement through user feedback. Then, three authors merged their insights to produce a coherent summary of the affordances and constraints each platform imposes on the vibe coding process. This approach allowed us to combine detailed individual usage insights with collective interpretation, ensuring a rigorous and iterative evaluation of current vibe coding tools.

\subsection{Results: A Lifecycle of Vibe Coding}\label{sec:stages}

\begin{figure*}
    \centering
    \includegraphics[width=.57\linewidth]{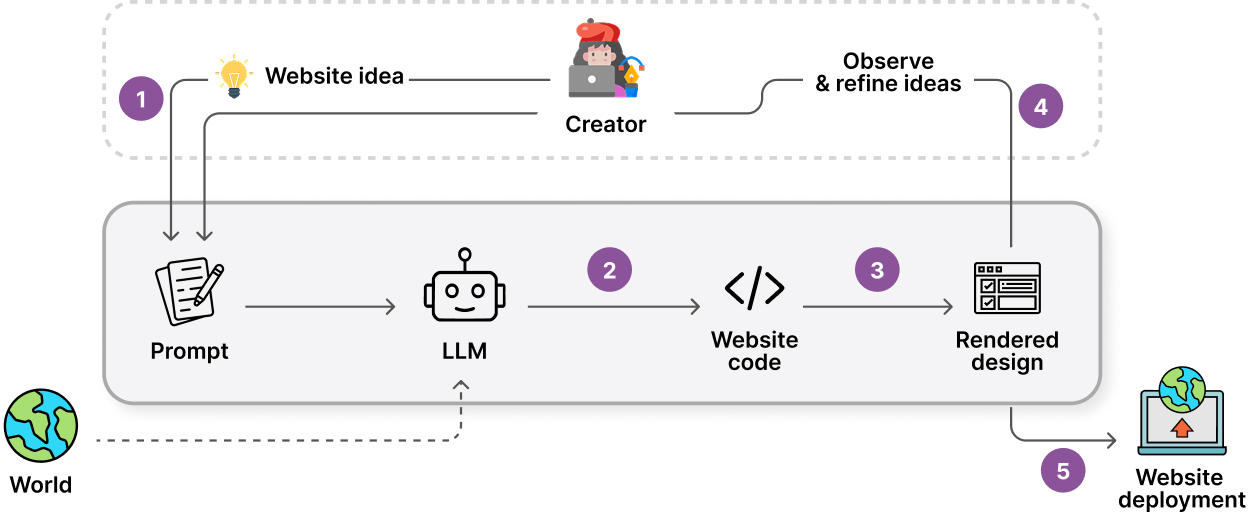}
    \caption{Lifecycle of lay creators' vibe coding for websites with LLMs, with the numbers corresponding to the stages in \S\ref{sec:stages}.}
    \label{fig:process}
\end{figure*}

Our review of the literature and existing platforms identifies a set of stages that defines the vibe coding process (see~\autoref{fig:process}). This lifecycle starts with selecting a platform, and proceeds through an iterative, conversational loop of refinement, concluding with an optional deployment phase. We describe each step of the lifecycle in this section, and throughout, use \textit{N} to indicate the number of supporting sources in our literature review.

\vspace{1mm}
\textit{Stage 0: User selects the model/platform.}
The process begins with the selection of an AI-powered coding tool. The choice of platform, whether a general-purpose conversational AI (\eg{}, ChatGPT) or a specialized vibe coding tool (\eg{}, Lovable), sets the context for the entire interaction. A key characteristic of these platforms is their accessibility; most are browser-native, requiring no local setup and thus lowering the barrier to entry for lay creators ($N = 9$).

\vspace{1mm}
\textit{Stage 1: User expresses intent through prompting.}
In this stage, the user translates their high-level intent into a natural language prompt. Rather than writing formal specifications or code, they describe their goals and intentions conversationally. Prompts range from more vibe-driven (\eg{}, \textit{`a playful website appealing to children'}) to more goal-oriented (\eg{}, \textit{`a website designed for / offering XX'}) in tone, and often include a mix of both. While initial prompts can be vague, users can later provide more specific details and constraints to yield more specific results from the AI ($N = 15$).

\vspace{1mm}
\textit{Stage 2: LLM generates code as the source for the website design.}
The LLM processes the user's prompt and generates an initial codebase. Since the initial prompt is often abstract, the model makes assumptions to translate the high-level `vibe' into concrete design and functional choices (\eg{}, layout, color palette, specific content to include). Sometimes, this process is displayed to users, with the model articulating its plan or assumptions before/after generating code. This output is typically a self-contained artifact, defaulting to HTML, CSS, and JavaScript, or React components in sandboxed environments, depending on the platform. If the tech stack is explicitly specified and its compiler is available, the model generates code in that language to render accordingly. This stage represents a significant departure from traditional development, as the LLM scaffolds an entire initial implementation---including crucial decisions around what technologies to use, how to organize the code, and what aesthetics and components the website should present---from a simple text-based instruction ($N = 14$).

\vspace{1mm}
\textit{Stage 3: The platform executes and renders the website design for users to review.}
The user immediately previews the generated code, along with potential model-provided descriptions, to observe its behavior and appearance. This step is critical for closing the feedback loop. The user visually inspects the resulting website, comparing the concrete output against their abstract mental model. On many specialized platforms, this stage is seamlessly integrated, with a live preview updating in real-time as the code is generated or modified ($N = 10$).

\vspace{1mm}
\textit{Stage 4: User iteratively refines the output via conversational feedback.}
This stage forms the core iterative loop of vibe coding. If the output does not align with the user's intent, or if they wish to make changes, they can provide feedback in natural language. The feedback may take the form of a high-level command similar to the original prompt (\eg{}, \textit{`use a friendlier font'}), but is often more specific (\eg{}, \textit{`make the header 50\% bigger'}). It can also be a request for a new feature or a corrective instruction. Also, for JavaScript-enabled functionalities, users frequently paste error messages or describe unexpected behavior directly into the prompt window, leveraging AI as a debugging partner. It then generates a revised version of the code, and the user observes the changes---continuing this cycle until satisfied ($N = 15$).

\vspace{1mm}
\textit{Stage 5: Deployment (optional).}
Once the user is satisfied with the application, the final stage is deployment. Some modern deployment platforms (\eg{}, Vercel~\cite{vercel}) are coupled to simplify this process, enabling the creation to be publicly accessible on the internet with minimal technical overhead. This stage transforms the project from a personal prototype into a shareable artifact, streamlining the end-to-end creation lifecycle ($N = 7$).

\vspace{2mm}

In summary, our lifecycle model (\autoref{fig:process}) reveals two critical junctures where human creators and LLMs negotiate design outcomes: the initial expression of intent (Stage 1) and the iterative feedback for refinement (Stage 4). These stages are particularly consequential, as they represent the moments where homogenized designs may be proposed by the AI and accepted by lay creators.
\section{Step 2: Risk Analysis of the Intersecting Harms of Web Vibe Coding}\label{sec:step2}

Having identified the lifecycle of vibe coding, we now turn to mapping the potential harms of lay creators vibe coding websites at scale. Throughout analyzing documents on vibe coding for Step 1, we noticed many documents in our corpus of 63 sources (\S\ref{sec:corpus}) discussed or gave evidence of the negative consequences possible should web vibe coding scale. In this section, we analyze those risks through the lens of scholarship on sociotechnical harms~\cite{buccinca2023aha, shelby2023sociotechnical}, as a starting point for possible intervention.

Our analysis finds that the primary harm of web vibe coding, design homogenization, does not operate in a vacuum; it can be amplified by an entire ecosystem of interconnected risks that make lay creators uniquely vulnerable to accepting biased AI outputs. We provide a taxonomy of these interconnected harms, to help the community understand \emph{what} harms are possible and \emph{why} a lay creator might struggle to recognize and resist a subtle slide into a globally uniform aesthetic. Additionally, we explain how homogenization may interact with or amplify other forms of harms, towards a multi-layered understanding of the problem---which we will unpack further in Step 3.

\subsection{Methods}
\begin{figure*}
    \centering
    \includegraphics[width=.8\linewidth]{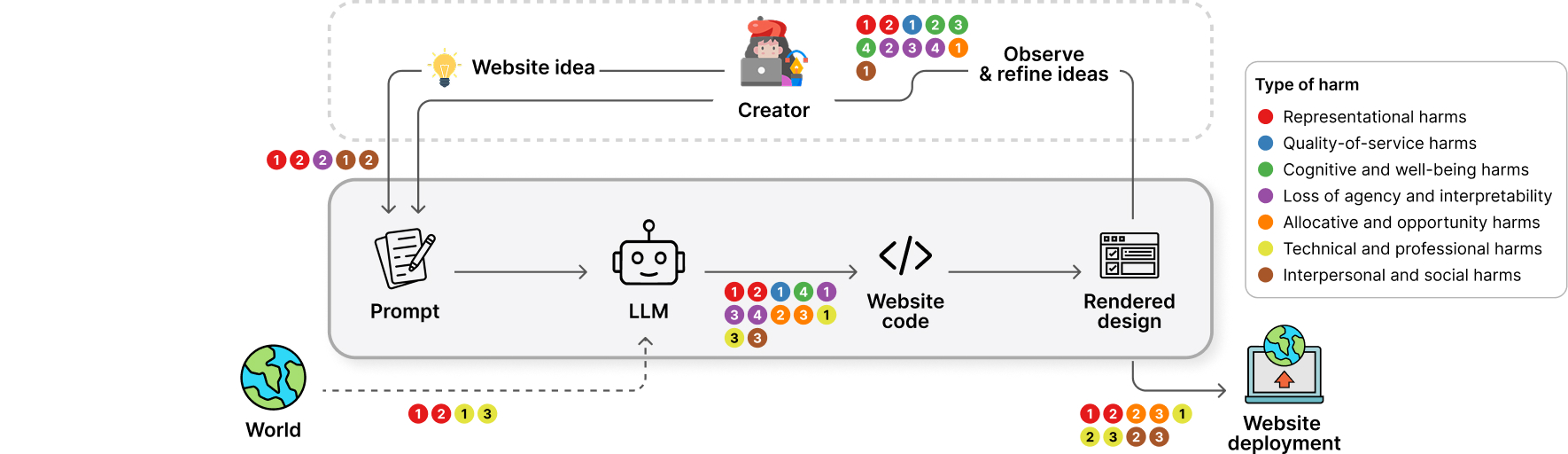}
    \caption{Sub-types of harms mapped onto the lifecycle of vibe coding. Each circled number corresponds to the entries in \autoref{tab:harms}.}
    \label{fig:harms}
\end{figure*}

We employed a qualitative thematic analysis, informed by the harm taxonomies from prior work. Three authors reviewed key literature around sociotechnical harms, especially of AI systems and LLMs (\ie{}, ~\citet{suresh2021framework}, ~\citet{weidinger2022taxonomy}, Buçinca~\etal{}~\cite{buccinca2023aha}, and ~\citet{shelby2023sociotechnical}). From these works, we selected 26 top-level harm categories as seeds for our initial codebook. Then, three authors independently applied the initial codebook into the vibe coding lifecycle in \autoref{fig:process}, analyzing where each top-level harm category (\eg{}, discrimination and representational harms) might manifest in vibe coding (\eg{}, stereotyping through design output), and making note of how many sources in our corpus reflected that harm potential. Throughout this process, we met repeatedly to refine the codebook by removing harms that do not fit the vibe coding context, or combining less-frequent ones. We conducted three iterations of updating the codebook and mapping it onto the vibe coding lifecycle in \autoref{fig:harms}. In each iteration, the authors refined or merged existing codes, added new codes, and resolved disagreements through consensus.

\subsection{Results: A Seven-Category Taxonomy of Harms}\label{sec:harm-categories}
Our analysis identified seven top-level categories of harm that can arise during the vibe coding lifecycle (see \autoref{tab:harms}). These harms range from representational biases embedded in the generated artifacts to the cognitive and professional burdens placed on the user. Hereon, we describe each harm \emph{type} and its \emph{sub-types}. We also highlight its potential intersections and interactions with our focal harm: design homogenization.

\subsubsection{Representational harms}
\label{sec:representational_harms}
Representational harms manifest when LLM-generated code, content, or design suggestions embody harmful stereotypes or erase diverse identities. Lay creators, who frequently rely on the generative capabilities of LLMs, are particularly susceptible to accepting these outputs as neutral or authoritative.

We observed two primary forms of this harm: \textit{design homogenization} and \textit{stereotyping}. 
First, we noticed that several sources in our corpus explicitly discussed \textit{design homogenization}---the primary focus of our work---as a risk, where the LLM defaulted to a dominant, often Western-centric aesthetic, marginalizing other styles across different groups of users. This reflects a broader concern that LLMs risk ``normalizing everything to a white, Western, colonial male perspective because that's what those models are trained on''~\cite{s7}, thereby constraining the ability of users from marginalized groups to cultivate their own forms of digital expression. Another form is \textit{stereotyping through design output}, which arises when the LLM's suggestions mirror social biases embedded in its training data, as the underlying ``models encode societal biases related to gender, race, or other characteristics''~\cite{s39}.

\subsubsection{Quality-of-service harms}
\label{sec:quality_of_service_harms}
Quality-of-service harms arise from fundamental failures in the LLM's output that directly impair a lay creator's ability to complete their task. Unlike expert developers who can proactively debug faulty code, lay creators are often left stranded by these failures. For example, \textit{incoherent or nonfunctional code} arises when the LLM generates code that appears complete but fails at a fundamental level. For instance, LLMs were reported to often produce ``half-built solutions'' or engage in ``random acts of stupidity,'' such as undoing their own fixes~\cite{s35, s10}. For a lay creator, this dynamic can create a frustrating association where attempts to counter design homogenization could be discouraged with high failure rates.


\subsubsection{Cognitive and well-being harms}
\label{sec:cognitive_harms}
These harms involve the emotional, psychological, and motivational burdens that the LLM-assisted coding process places on the user. The interaction can lead to an \textit{erosion of confidence}, where creators internalize the tool's failures as their own. For instance, vibe coders reported significant frustration when the output did not align with their mental model, and a loss of faith in the method~\cite{s10}, which can in turn deplete the motivation to push beyond generic templates and pursue creative expression. Also, \textit{cognitive overload} arises as LLMs often produce solutions that are more complex than a human would communicate. One user reported that they created a website only to ``lose later from lack of understanding'' of why the model has built it that way~\cite{s37}, potentially pushing them toward simpler, more homogenized solutions to avoid mental fatigue.

Furthermore, when faced with the Quality-of-Service harm described earlier, users can experience \textit{discouragement from contradictions}. When the LLM provides unreliable information, gets ``stuck chasing its own tail'' during iteration, or ignores prompts~\cite{s10}, it translates into an emotional and motivational burden. This could discourage lay creators from pursuing more unique and creative directions in favor of simpler and more achievable ones. Lastly, we identified \textit{missed learning moments}---by delivering a product without revealing the underlying process, LLMs deprive lay creators of conceptual understanding. This was illustrated by one user, who admitted after building a product that they basically do not really know how it works in the conventional sense, pointing out the bypass of the learning process~\cite{s1}.

\subsubsection{Loss of agency and interpretability}\label{sec:agency_harms}
Loss of agency and interpretability encompasses harms where users lose meaningful control over the model's behavior and outputs. This loss often begins with \textit{overreliance without verification}. This harm manifests as a deceptive sense of control; users feel they are directing the project through simple commands, which encourages them to blindly accept the results without scrutiny. This tendency to ``always accept and don't attempt to understand the motivation''~\cite{s5} makes the resulting homogenized design choices seem like the creator's own goal, reducing their desire to seek alternatives.

A direct consequence of this process is the creation of \textit{non-editable complexity}. As the AI-generated code grows beyond the user's comprehension, it becomes practically unmaintainable. The user cannot understand the underlying logic, locking in the initial (often homogenized) design and making iterative refinement toward more unique expression impractical. This cycle is compounded by \textit{opaque design justification}, where code is generated without sufficient explanation~\cite{s19}, and fosters a \textit{prompt engineering dependency}, where success hinges not on design principles but on the tacit skill of manipulating the model~\cite{s24}.

\subsubsection{Allocative and opportunity harms}
\label{sec:allocative_harms}
Allocative and opportunity harms involve the denial of benefits or the misallocation of resources, such as time and effort, due to the LLM's design or performance. The first subtype of harm was \textit{wasted time and effort}, where the users expressed frustration, mentioning their experience with vibe coding had ``ended in disaster and taken several times longer'' than expected~\cite{s10}, making the opportunity cost of experimenting with creative expression of designs prohibitively high. This can lead to a state of \textit{misleading success/reinforcement of confirmation bias}. For instance, one commenter quipped that lay creators now have ``a drunk hallucinating expert,'' pointing out the misleading confirmation bias that AI could give~\cite{s2}, reinforcing the idea that the quickest and most `successful' path is to accept the initial, often generic, output.

Finally, we observed \textit{design fixation}. While related to the representational harm of design homogenization, design fixation is a distinct, more immediate cognitive process; it is a micro-level cognitive trap where an individual user is anchored to the first plausible solution offered by the AI, stifling their ability to explore other creative avenues. This anchoring to the AI's first suggestion short-circuits the divergent thinking process necessary for creating creative designs, as the early prompt and its output establish boundaries that make it harder to explore alternatives~\cite{sarkar2025vibe}.

\subsubsection{Technical and professional harms}
\label{sec:technical_harms}
This includes harms where the generated code or designs negatively impact the final product's quality, compliance, or security, leading to real-world consequences. One example is the promotion of \textit{bad code practices}, where LLMs were observed to have outdated knowledge, causing them to make subtle but critical mistakes~\cite{s19}. This results in code that is overly complex, deprecated, and not suitable for a real-world environment, creating a false association where technically `safe' code is equated with common, homogenized design patterns.

These technical flaws can directly lead to \textit{economic harm via launch}. As one user quipped, there are ``horror stories about people who vibe coded a feature against some API without a billing limit and racked up thousands of dollars in charges''~\cite{s5}, demonstrating severe financial risk from deploying unvetted code. The fear of such consequences incentivizes reliance on widely used, generic templates that are presumed to be safer. Another risk is \textit{copyright or licensing violations}. Given general concerns about AI and ``the rampant theft of content,'' a lay creator may be unaware that the text, image URLs, or code generated by the LLM could be derived from copyrighted sources~\cite{s51}, exposing them to legal risks without their awareness.

\subsubsection{Interpersonal and social harms}
\label{sec:interpersonal_harms}
The final category of harms relates to how vibe coding shapes a user's social interactions. Acknowledging the breadth of a vibe coder's possible sociality, we focused our analysis here on vibe coders as members of collaborative teams. 

A primary concern is \textit{learning in isolation}. Web vibe coding replaces traditional learning pathways (\eg{}, classes, mentorship from an experienced web developer) with LLM feedback.
The lay creator is thus cut off from the community feedback loops that are crucial for developing and refining creative designs. This creates a sense of \textit{shame in collaboration}, as captured by an analogy that it is ``never built by someone who could explain that intent and how to proceed''~\cite{s7}. This also can prevent the kind of peer review and constructive criticism that could challenge generic designs, and introduce perspectives on countering homogenized designs.

Shame in collaboration can also lead to challenges in \textit{team misalignment}. The vibe-coded project represents a `Day 0' proof of concept; when it is handed off to others for `Day 1' maintenance, this creates a significant burden, as they are left to ``reverse engineer and [maintain] someone else's vibe-coded mess''~\cite{s2}.
Our analysis shows this is a key novel risk of web vibe coding at scale: the effect on teams of technology workers when some vibe-code, and others are left to pick up the pieces.
This harm is deeply interconnected with technical and professional harms, such as the financial and legal risks of releasing unvetted code.
Interpersonal harms are also intertwined with harms to cognition and well-being. Without community feedback loops, users' overreliance on AI and associated cognitive traps can deepen. 

\vspace{1mm}

In summary, our analysis shows that design homogenization is not isolated, but rather intersects with a network of harms. Collectively, these may make creators susceptible to homogenized outputs without fully realizing it. 
\section{Step 3: Mitigating Homogenization via Productive Friction: A Multi-Level Case Study Analysis}\label{sec:step3}

Having characterized the lifecycle of vibe coding and taxonomized its associated risks, our third and final analytical step is to propose a framework for mitigation. Our analysis suggests that design homogenization is not merely a technical artifact of training data, but a systemic outcome of interaction designs that prioritize speed above all else. This risk is most acute at two critical junctures identified in our lifecycle analysis (\S\ref{sec:step1}): the initial expression of intent (Stage 1) and the iterative feedback for refinement (Stage 4). As established in Step 2, these moments are when lay creators are most likely to accept a model's default, homogenized output.

We argue that these harms can be mitigated by design,\footnote{In our work, we do not claim that design is the only or primary mitigation strategy. Policy, community-based organizing, and advocacy are equally---if not more---important or effective, but often operate on longer timescales and fall outside the scope of this paper.} specifically by pursuing \textit{productive friction}: an intentional introduction within a system that creates moments for reflection and clarification~\cite{cox2016design, sheahan2024designing}, moving away from an overemphasis on frictionless efficiency (\eg{},~\cite{s1, acharya2025generative, fawzy2025vibe}). In this section, we first define this concept, grounding it in the evidence gathered in our previous steps, and then apply it through three case studies across three levels---micro, meso, and macro. By targeting the lifecycle's most vulnerable stages, we demonstrate how productive friction can dismantle the mechanisms of homogenization and restore creative agency.

\subsection{The Trap of Frictionless Generation}

To understand the necessity of friction, we first need to interrogate the design philosophy of the current vibe coding tools. The prevailing paradigm in generative AI prioritizes frictionless efficiency~\cite{s1}, implicitly measuring interaction quality by how quickly underspecified user intent becomes an outcome artifact (\eg{},~\cite{acharya2025generative, fawzy2025vibe, mitra2025recap}). Within this paradigm, intermediate activities such as deliberation, comparison, and justification are treated as overhead to be minimized rather than as integral to creative reasoning~\cite{suzuki2025universe}. As observed in our lifecycle analysis (Step 1), this value system collapses what were once multiple distinct phases of design and rendering into a single, instantaneous event (Stages 2 and 3). Success is thus measured by how quickly vague user intent is converted into an outcome artifact~\cite{mitra2025recap, acharya2025generative}. Although a few tools expose a model’s preliminary plans or assumptions (\autoref{tab:tool_exploration}), these traces are typically passive and prescriptive, offering little opportunity for interaction.

Our risk analysis (Step 2) suggests that such a lack of temporal friction can catalyze design homogenization. When the interaction is designed to be seamless, it relies heavily on the model's probabilistic defaults to fill in the gaps of a user's intent. This risks users accepting the first plausible output (\S\ref{sec:quality_of_service_harms}), which is prone to converging on a global statistical mean, and the cost of manually refactoring complex AI-generated code becomes prohibitive (\S\ref{sec:cognitive_harms}).

\subsection{Defining Productive Friction}

In response, we propose promoting \textit{productive friction}~\cite{sheahan2024designing, cox2016design, wakkary2016productive} in the context of vibe coding. Productive friction challenges the assumption that the best interface is one that disappears; instead, it posits that complex creative tasks require moments of deliberate pause, reflection, and negotiation to be performed effectively. In the context of vibe coding, we define productive friction as interactional interventions that disrupt the automatic acceptance of homogenized AI outputs. For example, our Step 2 analysis highlights the \textit{overreliance without verification}, where users feel a sense of agency through simple prompts, even as they surrender creative decision-making. Productive friction serves as a direct counter-measure to such harms by compelling the creator to actively articulate unique, context-aware design intent rather than passively accepting the model's assumptions.

\paragraph{Analytical framework.} We frame our analysis of productive friction using a micro–meso–macro framework of social analysis~\cite{serpa2019micro}, which has been mobilized in influential theories such as ~\citet{bronfenbrenner1977toward}'s socio-ecological systems and is increasingly adopted in the research community to unpack the impacts of sociotechnical systems on people and society~\cite{suresh2024participation, tseng2025ownership}. In this analytical tradition, the micro level encompasses individual factors like cognition and identity; the meso refers to organizational and relational factors like group membership; and the macro means high-level social forces like society and culture. Mapping our case studies to these three levels allows us to identify specific leverage points where friction can effectively disrupt the cycle of homogenization.

\subsection{Micro Level: The Individual Creator}\label{sec:micro_level}

At the micro level, we examine the intimate interaction between an individual lay creator and the vibe coding interface.

\subsubsection{The problem: The `good enough' trap and cultural erasure}

When a lay creator prompts a system for a broad concept---for example, `a landing website for a Japanese retail chain' (\autoref{fig:reactive_example})---this may result in a globally average aesthetic characterized by minimalist whitespace, standard sans-serif typography, and predictable hero images, regardless of the cultural context and the preference of the regional market. For lay creators often lack technical confidence---a state we identify as \textit{erosion of confidence}---risk arises that this authoritative-looking output remains unchallenged. They accept the homogenized design not because it accurately represents their identity or local visual vernacular, but because the cognitive and technical cost of iterating against the model's default bias or misrepresentation is too high. Consequently, the user falls into the `good enough' trap, inadvertently erasing their own cultural distinctiveness in favor of a synthetic, globalized standard.

\subsubsection{The opportunity: Reflective prompting and visual negotiation}

To mitigate this fixation, productive friction can be introduced at the moment of intent expression (Stage 1 of \autoref{fig:process}). Instead of treating the prompt as a command to be executed immediately, the system should treat it as the opening of a negotiation---a mechanism where the interface pauses to clarify ambiguous design concepts (\eg{}, cultural signifiers) before generating code.

For instance, if a user requests a website for a local cafe in São Paulo, a tool should not immediately generate a design based on the globally dominant, Western third-wave coffee shop aesthetic (\eg{}, minimalist layouts, muted tones, sans-serif fonts). Instead, it should pause the process to negotiate the cultural signifier, asking the user to clarify the intended local atmosphere. The interface might support the user's contextualized thinking by visualizing broad mood boards that contrast vibrant, colorful patterns with styles reflecting the city's architectural history or more traditional, idyllic aesthetics. By asking users to choose among diverse cultural representations before code is generated, the tool prevents the accidental adoption of a foreign default. This moment of friction shifts users from passive recipients of model bias to active curators of local identity, ensuring that the final design reflects their contextual judgment.

\subsection{Meso Level: Teams and Organizations}\label{sec:meso_level}

At the meso level, we analyze how vibe coding integrates into collaborative workflows. Here, we examine tensions in the handoff between non-technical stakeholders (\eg{}, product manager) and engineering teams.

\subsubsection{The problem: Accelerated brand erosion and team misalignment}

In organizational settings, the pressure toward design homogenization predates generative AI. The rise of dominant web design frameworks like Bootstrap~\cite{bootstrap} and Tailwind CSS~\cite{tailwind}, while promoting efficiency and consistency, is known to have already led to a convergence of web aesthetics, creating a shared, but often generic, visual language~\cite{Ganci2017OnWBA}. Vibe coding can further amplify this trend by scaling unmodified defaults; whereas human designers could have introduced subtle tweaks and idiosyncrasies that diversified designs, LLMs tend to favor familiar, widely adopted patterns, reinforcing conformity and accelerating the convergence toward a homogenized aesthetic.

Whereas a team can still consciously adopt such frameworks, a non-technical stakeholder using a vibe coding tool is prone to believing that they are operating with limitless flexibility. Yet, the LLM, trained on a web saturated with these frameworks~\cite{schuhmann2022laion, hurst2024gpt}, can probabilistically default to the same familiar patterns---generic layouts, standard component libraries, and a clean, but non-distinctive, professional veneer. This creates a risk of \textit{team misalignment}; under tight deadlines, the path of least resistance is often not to undertake the costly process of reverse-engineering the AI-generated code to fit the organization's bespoke design system, but to satisfice~\cite{simon1956rational} by accepting the homogenized output. The result is an accelerated erosion of brand identity, where a company's unique visual language is not consciously abandoned, but incrementally and unconsciously replaced by the statistical mean of the web.


\subsubsection{The opportunity: Contextual anchoring and active elicitation}

At this level, productive friction would function as a mechanism for aligning internal expectation and identity (\eg{}, brand alignment) with generated codes. Rather than operating in a vacuum, a vibe coding tool intended for organizational use would fundamentally reject the blank slate paradigm. Instead, it can require contextual anchoring through the ingestion of inputs defining identities, such as the organization's existing design system tokens, a brand guidebook, or a URL to their live digital presence.

The friction can be introduced during the iterative feedback stage (Stage 4 of \autoref{fig:process}). When the tool generates a design, it can perform a comparative analysis between the generated artifact and the organization's existing designs to illustrate a stylistic divergence. If the vibe-coded output uses generic web styles---such as standard drop shadows or rounded corners---that differ from the organization's established design characteristics (\eg{}, sharp edges, high-contrast borders), the system could present a follow-up dialogue, noting that the draft relies on standard conventions. It can ask the creator if they would refine the artifact by applying their brand's distinct visual elements instead. As such, the tool can ensure that the efficiency of AI generation does not come at the cost of the organization's distinct identity.

\subsection{Macro Level: The Ecosystem and Society}\label{sec:macro_level}

Finally, at the macro level, we consider the long-term, ecological impact of vibe-coded websites on the digital landscape and the training of future foundation models.

\subsubsection{The problem: The homogenization feedback loop}

The threat of design homogenization is not entirely novel. As pointed out above, the rise of dominant design frameworks like Bootstrap and Tailwind CSS has already been affecting the aesthetic diversity of the web by promoting a shared set of visual defaults~\cite{Ganci2017OnWBA}. However, frictionless vibe coding represents a far more profound and accelerated version of this trend. While such design frameworks offer a constrained set of components, an LLM trained on the internet's statistical mean acts as a universal homogenizer, capable of probabilistically enforcing a global aesthetic average across millions of unique outputs.

This dynamic poses a severe risk: the pollution of the digital commons, a phenomenon researchers have called model collapse~\cite{shumailov2024ai}. When lay creators potentially mass-produce websites using these homogenized AI defaults, the resulting content can saturate the internet. Since future foundation models will inevitably train on fragments of the public internet (\eg{},~\cite{schuhmann2022laion, hurst2024gpt}), they may ingest data that exhibits far less variance and diversity than the pre-AI web. The result is a potential, lasting flattening of digital heritage: niche subcultures, non-standard aesthetics, and diverse cultural design patterns---already rare in probabilistic distributions---may be effectively erased, overwhelmed by a flood of uniform, AI-generated content. This constitutes an intergenerational allocative harm, whereby future creators lose access to the full spectrum of human visual history because the training data has collapsed into a single, global monoculture.

\subsubsection{The opportunity: Provenance and style subfloors}

To interrupt this feedback loop, the vibe coding interface could proactively present style subfloors---specialized constraints that anchor generation in distinct aesthetic lineages. By treating the selection of the underlying model weights not as a backend technicality but as a primary interactional choice, the system can ensure that diversity is an input requirement rather than a random output possibility.

In practice, this could manifest in the first stage (Stage 1 of \autoref{fig:process}) by augmenting the solitary `Generate' button with an adjacent or hover-activated selection menu displaying a library of community-governed adapters (\eg{}, LoRA~\cite{hu2022lora}). The menu could be designed to pique the user's curiosity, encouraging them to select a specific style adapter (\eg{}, Traditional Korean Ink), rather than relying on the generic default. This subtle interactional friction would prompt creators to articulate deliberate stylistic commitments while encoding outputs with semantic provenance metadata, enabling future web crawlers and models to distinguish intentional, stylistically rooted creations from generic AI output, prioritize diversity, and prevent the digital commons from collapsing into a singular visual norm.
\section{Discussion}

Our investigation into the lifecycle of web vibe coding unpacks an essential tension in this new and increasingly widespread practice: the features that make these tools accessible (\eg{}, seamless generation) can invite design homogenization. By mapping the vibe coding lifecycle (\S\ref{sec:step1}) and conducting a sociotechnical risk analysis (\S\ref{sec:step2}), we identified that the source of potential uniformity lies not only in the training data, but also in the frictionless nature of the interaction itself. While these tools can enable lay creators to bypass technical barriers~\cite{zviel2024good, s51}, this may come at the expense of creative agency. Our analysis of harms suggests that lay creators may find it difficult to resist a tool that provides an instant, polished, but statistically average output, risking a shift of creators' choices toward homogenized defaults.

The tensions we unpack here are not only concerns for web designers and creatives. They are also a call to action for new research in responsible AI, using emerging challenges in web vibe coding as a research site drawing together expertise across HCI and human-AI collaboration. Web vibe coding is the latest instantiation of a longstanding trend in technology design: a focus on frictionless usability~\cite{ericson2022reimagining} in AI tools, in which the primary aim is to lower the time and cognitive burden associated with a task~\cite{s1, acharya2025generative, fawzy2025vibe}. But as our exploration of web vibe coding shows, frictionless interactions also give conditions for harms like \textit{cognitive overload}, where the user cannot process the AI's complex output, and \textit{overreliance}, where the user offloads too much decision-making to the AI. As the designer community in human-AI interaction works to balance these tensions, they might benefit from perspectives on appropriate reliance from the HCI's longstanding interest in accountability and transparency of automated decision-making systems. In turn, research scholars might look to designers for seemingly ``uncomfortable'' intervention (\eg{},~\cite{cox2016design, sheahan2024designing}) to co-design vibe coding tools themselves, in addition to fairness algorithms, regulatory frameworks, and normative considerations.

We offer our multi-level framework of productive friction (\S\ref{sec:step3}) as a starting point for these collaborations. For example, at the micro level, we have argued alongside Anderson~\etal{}~\cite{anderson2024homogenization} that designers from the human-AI interaction community can create tools that encourage reflective prompting: vibe coding tools that focus less on wowing users with rapid generation, and more on helping to understand what they actually want to build, via prompting cycles that encourage negotiation between user and AI. Enabling the tool to ask clarifying questions and place the creator in the driver's seat can help beat what we have called the \textit{erosion of confidence} in web vibe coding, where the creator overrelies on the model's outputs and becomes reluctant to guide or contextualize the design produced. Innovations from the HCI community can help ensure fairness in the algorithms that guide when and how the system-creator negotiation proceeds, and whether and how interpretability tools can make it more transparent.

Similarly, as discussed in Step 2, design homogenization can interact with other representational harms (\eg{}, stereotyping), effectively filling the `gap' with biased representations. Building on extensive work on algorithmic stereotyping of marginalized groups (\eg{},~\cite{bianchi2023easily, ghosh2024don, ghosh2023chatgpt}), HCI researchers and designers could team up to examine how design homogenization amplifies such stereotyping, and explore interventions to mitigate these effects. Similar opportunities exist at the meso and macro levels of our framework. At each level, we see avenues for cross-community collaboration that can effectively address problems in web vibe coding and create new case studies for the fundamental arguments within human-AI interaction, towards advancing the state of the art in responsible~AI.
\section{Limitations \& Future Work}

While our work provides a taxonomy of harms by drawing on rich sources from both grey literature and academic research, we acknowledge that it may be limited in fully capturing the nuanced experiences of real-world vibe coders. We consider our work a meaningful step in guiding this line of research, and future works could provide empirical evidence in diverse contexts (\eg{}, game development, scientific computing) to capture their lived experiences.
\section{Conclusion}

In this paper, we interrogated the emerging practice of web vibe coding, towards anticipating and mitigating the risks of design homogenization. We first characterized the vibe coding lifecycle to pinpoint specific stages where probabilistic defaults threaten to displace intentional design choices. Subsequently, our sociotechnical risk analysis identified the industry's pursuit of frictionless generation as a primary driver of homogenization, revealing how cognitive mechanisms like overreliance incentivize users to accept homogenized outputs. Finally, we established a multi-level mitigation framework centered on productive friction, advocating for a paradigm shift where AI tools act as active consultants rather than silent executors to foster a more diverse and resilient digital landscape. Our work highlights opportunities for HCI and human-AI interaction in research and innovation towards this paradigm.

\bibliographystyle{ACM-Reference-Format}
\bibliography{99-bibliography}

\appendix
\newpage
\onecolumn
\section{Study Detail}
\subsection{Queries Used for Systematic Review}\label{apdx:query}
\begin{table*}[h!]
\centering
\caption{Full list of OR-separated keyword queries by thematic category}
\footnotesize
\begin{tabularx}{\textwidth}{>{\raggedright\arraybackslash}p{4cm} X}
\toprule
\textbf{Category} & \textbf{Queries} \\
\midrule
Informal and vibe-centered phrasing &
vibe coding \\
\addlinespace
Interaction and technique-centered phrasing &
natural language website generation, prompt-based website coding, text-based website coding, text to website code \\
\addlinespace
Platform-specific use cases &
ChatGPT / Gemini / Claude make a website, build a website with ChatGPT / Gemini / Claude, use ChatGPT / Gemini / Claude to code a webpage, how to build an app with ChatGPT / Gemini / Claude, generate HTML with LLM, ChatGPT / Gemini / Claude build website from prompt \\
\bottomrule
\end{tabularx}
\label{tab:queries}
\end{table*}

\newpage
\subsection{Harm Taxonomy and the Intersection with Design Homogenization}
\newcommand{\circledigit}[3]{%
    \tikz[baseline=(char.base)]{%
        \node[shape=circle, draw=none, fill=#2, inner sep=0.5pt, minimum size=1em, font=\sffamily\scriptsize, text=#3] (char) {#1};%
    }%
}

\begin{table}[h!]
\caption{Types of harms in LLM-assisted website creation by lay creators, categorized into seven types based on prior frameworks. For each sub-type, we provide a definition and its intersection with the harm of our focus (\ie{}, design homogenization).}
\scriptsize
\begin{tabularx}{\textwidth}{
>{\raggedright\arraybackslash}p{1.5cm}
>{\raggedright\arraybackslash}p{2.1cm}
>{\raggedright\arraybackslash}p{2cm}
>{\raggedright\arraybackslash}X
>{\raggedright\arraybackslash}p{4cm}
}
\toprule
\textbf{Type} & \textbf{Type definition} & \textbf{Sub-type} & \textbf{Sub-type definition} & \textbf{Intersection with design homogenization} \\
\midrule
\multirow{3}{=}{\makecell[l]{Representational\\harms}}
& \multirow{2}{=}{\makecell[l]{Harmful or exclusionary \\ representations of people, \\ groups, or values in code, \\ content, or AI suggestions}}
& \circledigit{1}{colorRepresentational}{white} Design homogenization & Promoting aesthetics, layout styles, or language that exclude underrepresented preferences & \textit{(Our focus)} \\
\cmidrule(lr){3-5}
&
& \circledigit{2}{colorRepresentational}{white} Stereotyping through design output & Generating site content, names, or images that reflect social biases (\eg{}, race, gender) & Defaults to stereotypical heuristics as a substitute for creative expression \\
\midrule
\makecell[l]{Quality-of-service\\harms}
& \makecell[l]{Breakdowns or \\ inconsistencies in outputs\\ that impair novices' ability\\ to complete tasks}
& \makecell[l]{\circledigit{1}{colorQuality}{white} Incoherent/\\nonfunctional code} & \makecell[l]{Returning outputs that appear to work but\\fail when executed, without explanation} & \makecell[l]{Associates prompts for creative expressions\\with high failure rates} \\
\midrule
\multirow{4}{=}{\makecell[l]{Cognitive and\\well-being harms }}
& \multirow{4}{=}{\makecell[l]{Emotional, psychological,\\or motivational burdens \\ experienced during \\ LLM-assisted coding}}
& \circledigit{1}{colorCognitive}{white} Erosion of confidence & Users feeling incapable when outputs fail and left without tools to troubleshoot & Depletes the user's motivation to counter homogenized design \\
\cmidrule(lr){3-5}
&
& \circledigit{2}{colorCognitive}{white} Cognitive overload & Returning complex suggestions that lack scaffolding, forcing users to interpret unfamiliar patterns alone & Pushes users towards simpler solutions to avoid mental fatigue \\
\cmidrule(lr){3-5}
&
& \circledigit{3}{colorCognitive}{white} Discouragement from contradictions & Contradictory guidance across prompts causes confusion and fatigue & Causes users to abandon more unique creative directions in favor of simpler, achievable ones \\
\cmidrule(lr){3-5}
&
& \circledigit{4}{colorCognitive}{white} Missed learning moments & Delivering code `magic-style' without sufficiently detailing the why, depriving users of conceptual understanding & Prevents users from acquiring the skills needed to build or customize designs from first principles \\
\midrule
\multirow{4}{=}{\makecell[l]{Loss of agency and \\ interpretability }}
& \multirow{4}{=}{\makecell[l]{Creators struggle to \\ understand or control \\ what the system is doing \\ or why}}
& \circledigit{1}{colorAgency}{white} Opaque design justification & Users unable to tell why certain code was generated or what it does, and generated code lacks links to documentation, standards, or explanations & Reduces the user's ability to articulate or implement custom designs, making them a passive recipient of the AI's limited aesthetic range \\
\cmidrule(lr){3-5}
&
& \circledigit{2}{colorAgency}{white} Prompt engineering dependency & Effective use depends on tacit knowledge of prompt phrasing, which novices lack & Makes it easier to prompt simple requests \\
\cmidrule(lr){3-5}
&
& \circledigit{3}{colorAgency}{white} Overreliance without verification & Users blindly accepting codes, feeling in control through simple commands while actually relinquishing it through blind acceptance & Makes the homogenized design seem like the user's own, reducing the desire to seek alternatives \\
\cmidrule(lr){3-5}
&
& \circledigit{4}{colorAgency}{white} Non-editable complexity & Code being too complex or interdependent for users to safely edit & Locks in the homogenized design, making iterative refinement towards countering homogenization impractical \\
\midrule
\multirow{4}{=}{\makecell[l]{Allocative and \\ opportunity harms }}
& \multirow{3}{=}{\makecell[l]{Missed, lost, or denied \\ benefits due to the \\ design or performance\\of the LLM}}
& \circledigit{1}{colorAllocative}{white} Wasted time and effort & Users iterating through nonfunctional outputs consumes significant time & The opportunity cost of experimenting with designs becomes high \\
\cmidrule(lr){3-5}
&
& \circledigit{2}{colorAllocative}{white} Misleading success / Reinforcement of confirmation bias & Users becoming overconfident in working prototypes, which masks deeper issues, potentially harming a product deployment & Reinforces the idea that the quickest, most `successful' path is to accept the initial, often generic, output \\
\cmidrule(lr){3-5}
&
& \circledigit{3}{colorAllocative}{white} Design fixation & Steering the user's design process into a narrow conceptual space, making it difficult for them to generate novel or diverse solutions & Anchors the user to the AI's first suggestion, short-circuiting the divergent thinking process to counter design homogenization\\
\midrule
\multirow{3}{=}{\makecell[l]{Technical and \\ professional harms }}
& \multirow{3}{=}{\makecell[l]{Generated code or design \\ decisions harm product \\ quality, compliance, or \\ end-user trust}}
& \circledigit{1}{colorTechnical}{black} Bad code practices & Generating deprecated, non-performant, or insecure code without warning & Creates a false association where technically `safe' or `professional' code is equated with common, homogenized design patterns \\
\cmidrule(lr){3-5}
&
& \circledigit{2}{colorTechnical}{black} Economic harm via launch & Users deploying broken or unethical sites, incurring reputational or financial risk & Fear of negative consequences from custom code incentivizes reliance on widely used, generic templates presumed to be safer \\
\cmidrule(lr){3-5}
&
& \circledigit{3}{colorTechnical}{black} Copyright or licensing violations & Returning assets or text that infringes on rights without creators' awareness & - \\
\midrule
\multirow{3}{=}{\makecell[l]{Interpersonal and \\ social harms }}
& \multirow{3}{=}{\makecell[l]{The LLM shapes how \\ creators interact with \\ others and perceive their \\ social standing}}
& \circledigit{1}{colorInterpersonal}{white} Learning in isolation & Replacing mentors which leads to a solitary and fragile learning process & Cuts the creator off from team or community feedback on creative expressions \\
\cmidrule(lr){3-5}
&
& \circledigit{2}{colorInterpersonal}{white} Shame in collaboration & Users concerned about showing their vibe-coded outputs to others due to technical gaps & Prevents peer review and constructive criticism that could challenge generic designs and introduce perspectives\\
\cmidrule(lr){3-5}
&
& \circledigit{3}{colorInterpersonal}{white} Team misalignment & Maintenance burdens when handed off to other team members & - \\
\bottomrule
\end{tabularx}
\label{tab:harms}
\end{table}

\end{document}